\documentclass{article}

\usepackage{microtype}
\usepackage{graphicx}
\usepackage{subfig}
\usepackage{booktabs} 
\usepackage{tabularx}
\usepackage{mathtools} 
\usepackage{bm}

\usepackage{hyperref}

\usepackage{amsmath}

\usepackage[accepted]{icml2024}

\usepackage{amsmath}
\usepackage{amssymb}
\usepackage{mathtools}
\usepackage{amsthm}

\usepackage[capitalize,noabbrev]{cleveref}

\theoremstyle{plain}

\theoremstyle{definition}

\theoremstyle{remark}

\usepackage[textsize=tiny]{todonotes}

\icmltitlerunning{A Physics-Informed Machine Learning Approach for Solar Irradiance Estimation}

\begin{document}

\twocolumn[
\icmltitle{A Physics-Informed Machine Learning Approach utilizing Multiband Satellite Data for Solar Irradiance Estimation}



\icmlsetsymbol{equal}{*}

\begin{icmlauthorlist}
\icmlauthor{Jun Sasaki}{equal,jwa}
\icmlauthor{Maki Okada}{equal,jwa}
\icmlauthor{Kenji Utsunomiya}{jwa}
\icmlauthor{Koji Yamaguchi}{jwa}
\end{icmlauthorlist}

\icmlaffiliation{jwa}{Department of Environment and Energy, Japan Weather Association, Tokyo, Japan}

\icmlcorrespondingauthor{Jun Sasaki}{sasaki.jun@jwa.or.jp}
\icmlcorrespondingauthor{Maki Okada}{okada.maki@jwa.or.jp}

\icmlkeywords{Machine Learning, ICML}

\vskip 0.3in
]



\printAffiliationsAndNotice{\icmlEqualContribution} 

\begin{abstract}
Solar irradiance is fundamental data crucial for analyses related to weather and climate.
High-precision estimation models are necessary to create areal data for solar irradiance.
In this study, we developed a novel estimation model by utilizing machine learning and multiband data from meteorological satellite observations.
Particularly under clear-sky and thin clouds, satellite observations can be influenced by surface reflections, which may lead to overfitting to ground observations.
To make the model applicable at any location, we constructed the model incorporating prior information such as radiative transfer models and clear-sky probability, based on physical and meteorological knowledge.
As a result, the estimation accuracy significantly improved at validation sites.
\end{abstract}

\section{Introduction}

Solar irradiance data is essential for monitoring solar power, assessing potential output, climate simulations, and climate monitoring.
These applications often require high-precision areal data reflecting regional characteristics.
In principle, it is possible to estimate solar irradiance through radiative transfer calculations \cite{mueller2009cm,xie2016fast,zhang2018estimation};
however there are some challenges such as cloud modeling and computational costs.
As a result, statistical models based on meteorological satellite data are widely used \cite{engerer2017himawari,saito2018,hashimoto2023development}.

However, these standard models are simplistic, utilizing limited band information.
For example, Saito et al. \yrcite{saito2018} uses visible and infrared band data for weather classification with rule-based algorithms and piecewise linear regression for estimation.
Actually, the relationship between solar irradiance and satellite observations is highly complex due to the diversity of clouds,
suggesting that combining multiband observation data with machine learning models can lead to more precise estimations of solar irradiance.




In previous studies where machine learning was applied to solar irradiance estimation models,
primarily limited elements such as the visible band and simpler architectures like MLPs (Multilayer Perceptrons) were used,
indicating that there is potential for improvement in accuracy \cite{cornejo2019machine,jiang2019deep,yeom2019spatial,palacios2022machine}.


This study aims to develop a high-precision model applicable at any location, leveraging multiband satellite images and machine learning.
To avoid overfitting to ground observations and ensure applicability at any site, we incorporated constraints based on physical and meteorological knowledge to enhance the model's generalizability.
we have quantitatively demonstrated the model's superiority by comparing it with the Saito's standard model \yrcite{saito2018} under consistent conditions.

\section{Data description}

In this study, we used satellite image data from Japan's Himawari-8, the imager on board of which has 16 observation bands ranging from \( 0.47\mu \)m to \( 13.3\mu \)m in wavelength.
Six of these bands are in the shortwave spectrum (visible and near-infrared) and ten in the longwave spectrum (infrared), with the visible bands strongly correlating with solar irradiance due to their high spectral intensity.
Near-infrared bands reflect the size and phase state (water or ice) of cloud particles, and infrared bands vary in vertical sensitivity according to their wavelengths, providing vertical structural information of clouds \cite{shimizu2017}.

For model training, we used one-minute interval ground observations from 47 sites across Japan, provided by the Japan Meteorological Agency.
For validation, we used solar irradiance data independently observed in this study \cite{ishii2013simplified}.
The observation sites are Naganuma, Tsukuba, Gifu, Tosu, and Okinoerabu, referred to as ST1, ST2, ST3, ST4, and ST5, respectively.
Each site has distinct solar irradiance characteristics and surface conditions, allowing for a comprehensive evaluation of the model's accuracy and generality across diverse environmental conditions.
The model was trained over four years, from 2016 to 2020, excluding 2018, with a specific test period set for 2018.


\section{Methodology}

In this study, we utilized a deep learning framework to model the relationship between satellite imagery and solar irradiance.
Specifically, we input data from 16 satellite observation bands and trained the model using ground-measured solar irradiance as the truth values.
However, under clear skies and thin clouds, the contribution of reflections and emissions from the surface to the satellite observations is significant.
Due to the limited number of ground observation sites compared to the diversity of surface conditions (such as urban areas, forests, seas, and snow covers),
there is a risk of the model overfitting to the characteristics of the observation sites.
To address this, we incorporated a radiative transfer model and clear-sky probability to improve the model's generalizability to unknown locations.
The developed model consists of the following three sub-models:

\begin{itemize}
\item Clear-Sky Solar Irradiance Estimation Model.
\item All-Sky Solar Irradiance Estimation Model.
\item Clear-Sky Probability Estimation Model.
\end{itemize}

\subsection{Clear-Sky Solar Irradiance Estimation Model}

This sub-model is based on the model by Kondo \yrcite{kondo1994} and is represented by the following equations:
\begin{align}
& I_{\text{clr}} = a_0 I_{0N} \cos Z (1 + a_1 \times 10^{-a_2 m}) C_w \label{clrsky_radiation} \\
& C_w = [1 + a_3 \log_{10} w + a_4 \log^2_{10} w + a_5 m \log_{10} w] \label{effect_water_vapor}
\end{align}
Where \( I_{\text{clr}} \) is the solar irradiance under clear-sky conditions, \( I_{0N} \) is the extraterrestrial solar irradiance,
\( Z \) is the solar zenith angle,
\( m \) is the air mass, and \( w \) is the precipitable water vapor.
\( C_w \) represents the term for seasonal variation of solar irradiance due to changes in water vapor amount.
The precipitable water vapor \( w \) is estimated based on the date and latitude of the target site.
The parameters \( a_0 \) to \( a_5 \) are optimized using ground-measured solar irradiance data.
Since this model is determined only by the location and date/time and does not rely on satellite observations, it can be applied to any region.

\subsection{All-Sky Solar Irradiance Estimation Model}

\begin{figure}[ht]
\vskip 0.1in
\begin{center}
\centerline{\includegraphics[trim=0cm 0cm 0cm 0cm, width=\columnwidth]{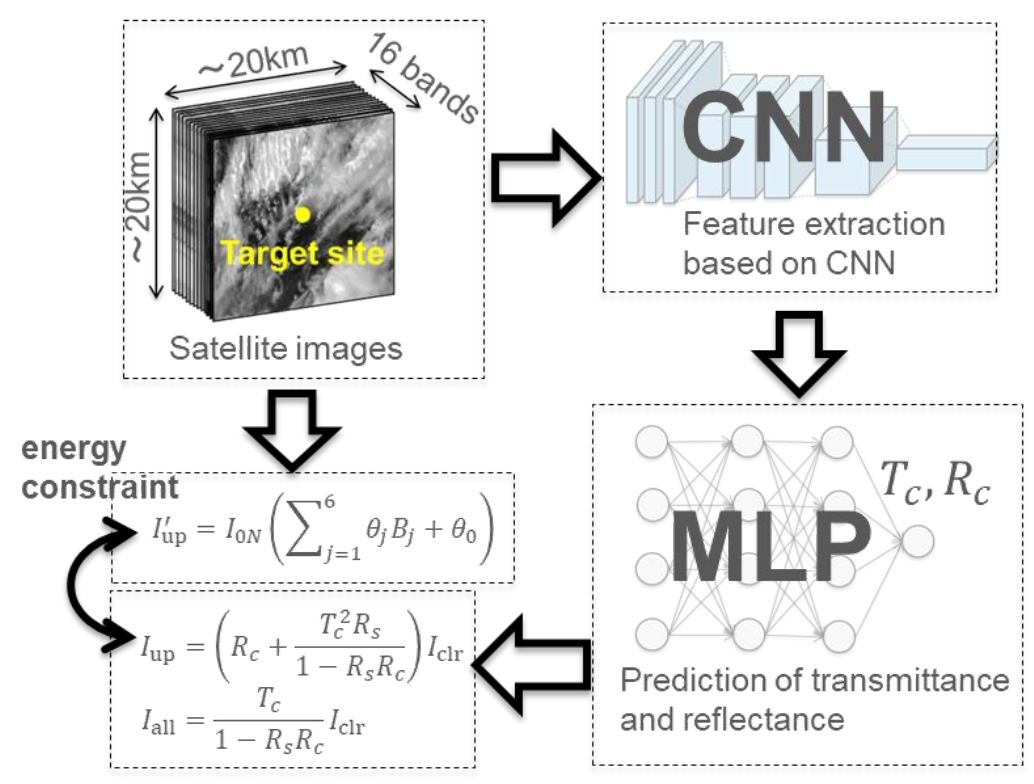}}
\caption{Architecture of the all-sky solar irradiance estimation model.
Satellite images from multiple bands are cropped around the target site and used as input.
Features of the spatial distribution of clouds are extracted using a CNN, and the transmittance and reflectance of clouds are predicted using an MLP.
Finally, downward and upward shortwave radiation are calculated using radiative transfer equations.
Additionally, upward shortwave radiation is directly calculated from satellite observation data.}
\label{allsky_estimation_model3}
\end{center}
\vskip -0.2in
\end{figure}

This sub-model calculates the all-sky solar irradiance from 16-band satellite images. First, we formulated a radiative transfer model based on Liou \yrcite{liou2002introduction}.
When the flux corresponding to clear-sky irradiance, \( I_{\text{clr}} \),
is incident on a multilayer atmospheric system consisting of clouds, atmosphere, and surface, the radiative transfer equations,
considering multiple reflections between clouds and the surface, are expressed as follows:
\begin{equation}
 I_{\text{all}} = \frac{T_c}{1 - R_s R_c} I_{\text{clr}}, I_{\text{up}} = \left(R_c + \frac{T_c^2 R_s}{1 - R_s R_c}\right) I_{\text{clr}}
    \label{eq:transfer_equation}
\end{equation}
Here, \( I_{\text{all}} \) represents the downward shortwave radiation at the surface, corresponding to the all-sky solar irradiance, and
\( I_{\text{up}} \) is the upward shortwave radiation at the top of the atmosphere.
\( T_c \) and \( R_c \) are the transmittance and reflectance of clouds, respectively, and \( R_s \) represents the surface albedo.
\( I_{\text{clr}} \) is estimated using equation \ref{clrsky_radiation},\ref{effect_water_vapor}, and \( R_s \) is estimated from the most recent visible band images under clear-sky conditions.

The transmittance \( T_c \) and reflectance \( R_c \) are calculated using a neural network with multiband satellite images as inputs.
The architecture, as shown in \cref{allsky_estimation_model3}, consists of a CNN (Convolutional Neural Network) and an MLP.
Here, the input data comprises satellite images of a rectangular area approximately 20 km around the target site.
It is expected that the CNN can extract features of the cloud's spatial structure, thereby improving the representation of scattered irradiance.

\begin{figure}[ht]
\vskip 0.1in
\begin{center}
\centerline{\includegraphics[trim=0cm 0cm 0cm 0cm,height=3cm]{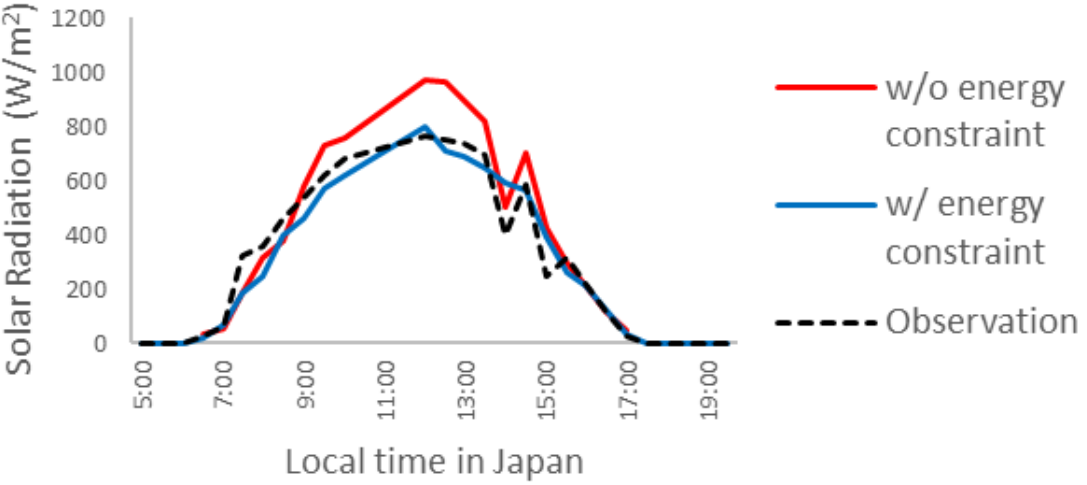}}
\caption{Differences in estimated solar irradiance with and without radiative energy constraints.
Black dotted, red, and blue lines indicate ground observations, estimation without constraints, and estimation with constraints, respectively.
Time series for ST1, March 7, 2018.}
\label{energy_constraint2}
\end{center}
\vskip -0.2in
\end{figure}

The calculated \( T_c \) and \( R_c \) are converted into the upward and downward shortwave radiations through equation \ref{eq:transfer_equation}.
Here, \( I_{\text{all}} \) is optimized using the ground-measured irradiance \( I_s^{(\text{o})} \),
while \( I_{\text{up}} \) corresponds to the shortwave radiation observed by the satellite, integrated over the wavelength region, denoted as \( I_{\text{up}}' \), which can approximately be expressed as:
\begin{equation*}
     I_{\text{up}}' = I_{0N} \left(\sum\nolimits_{j=1}^{6} \theta_j B_j + \theta_0 \right)
\end{equation*}
Here, \( B_j \) represents the shortwave radiation observed by the satellite, and \( \theta_j \) are the learning parameters.
Assuming that absorption by the atmosphere, ice clouds, and water clouds is low \cite{hale1973optical,warren1984optical},
 \( R_c + T_c \approx 1 \), the overall loss function can be expressed using the hyperparameters \( \alpha \), \( \beta \), \( \gamma \) as follows:
\begin{equation*}
 \mathcal{L} = \alpha \|I_{\text{all}} - I_s^{(\text{o})}\|_2 + \beta \|R_c + T_c - 1\|_2 + \gamma \|I_{\text{up}} - I_{\text{up}}'\|_2
\end{equation*}
Through optimization with this loss function, the energy balance is maintained so that \( I_{\text{all}} \) fits the ground observations while \( I_{\text{up}} \) remains consistent with satellite observations, \( I_{\text{up}}' \),
thus preventing overfitting to observation sites.
In practice, optimizing without energy-related constraints resulted in outliers under complex conditions such as thin clouds over snow cover (\cref{energy_constraint2}).



\subsection{Clear-Sky Probability Estimation Model}

\begin{figure}[ht]
\vskip 0.1in
\begin{center}
\centerline{\includegraphics[trim=0cm 0cm 0cm 0cm, height=4.5cm]{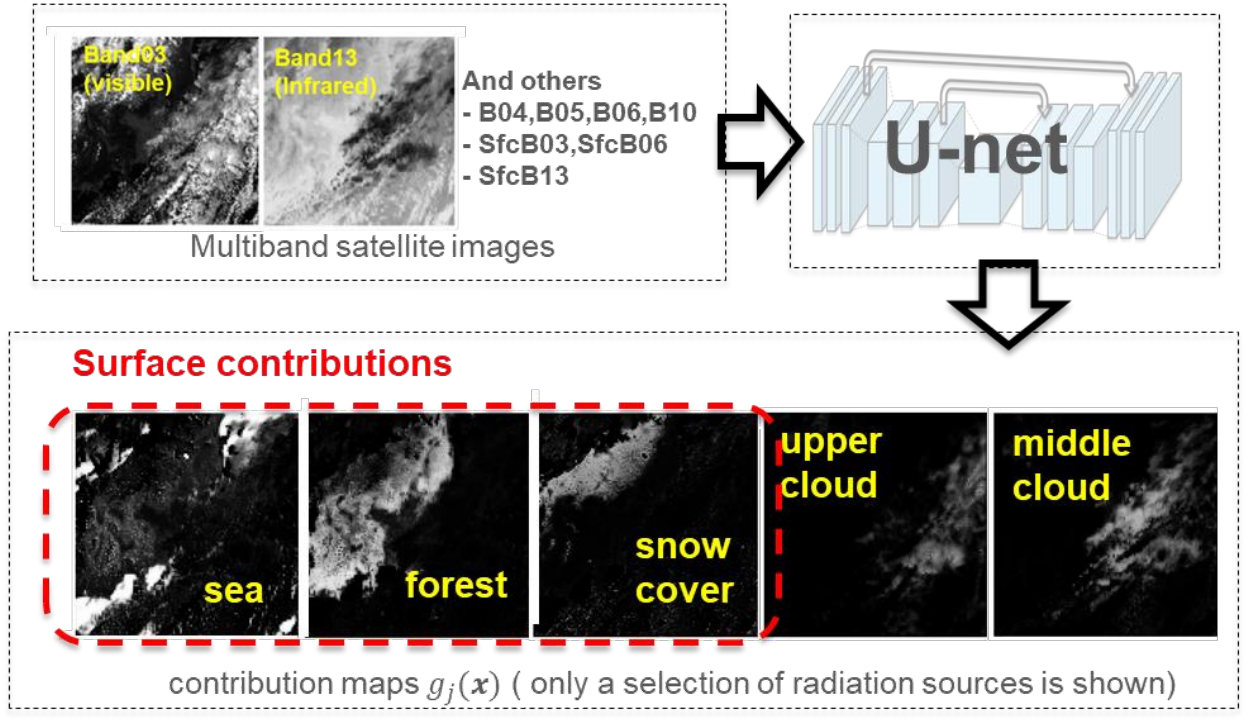}}
\caption{Processing flow and example results of the clear-sky probability estimation model.
Satellite images from multiple bands are used as inputs to U-net, which outputs the contribution maps for different radiation sources such as seas, forests, snow covers, upper clouds and middle clouds.
Input data include images from various bands (B03, B04, B05, B06, B10, B13) as well as surface albedo from Band03 and Band06 (SfcB03, SfcB06), and surface brightness temperature from Band13 (SfcB13).
The results shown here are for the Kanto region, spanning latitudes 35°N to 38°N and longitudes 137°E to 143°E, on January 15, 2018, at 14:00.}
\label{satellite_segmentation5}
\end{center}
\vskip -0.2in
\end{figure}

Especially under clear sky conditions, the influence of reflections and emissions from the surface is significant,
and there are limitations in enhancing the generalizability of the all-sky solar irradiance estimation model to unknown locations.
Therefore, we introduce the clear-sky probability, \( P_{\text{clr}} \), defined as a random variable that takes the value 1 in clear-sky and 0 in cloudy-sky conditions.
The final estimated solar irradiance, \( I_s \), is calculated as the expected value based on the clear-sky probability:
\begin{equation*}
 I_s = P_{\text{clr}} \times I_{\text{clr}} + (1 - P_{\text{clr}}) \times I_{\text{all}}
\end{equation*}
This approach reduces the influence of the surface and ensures accuracy for unknown locations by using a larger proportion of the clear-sky solar irradiance, \( I_{\text{clr}} \),
which is independent of satellite observations, under clear-sky conditions.


This sub-model is designed based on the model by Akimoto et al. \yrcite{akimoto2020fast}.
First, we perform cluster analysis on states defined by combinations of multiband satellite observation values, \( B_i \) (where \( i \) denotes the band type), such as \( (B_{i_1},B_{i_2},...) \).
The centroid of each cluster, \( b_{ji} \) (where  \( j \) is the cluster number), can be considered as representative states corresponding to different radiation sources.
By estimating coefficients \( g_j \) that reproduce the original satellite observation values \( B_i \) through a linear combination of \( b_{ji} \),
\( g_j \) can be considered as the contribution of each radiation source to the satellite observations.
In practice, a U-net \cite{ronneberger2015u} was optimized to output the contribution map \( g_j(\bm{x}) \) from the input satellite images \( B_i(\bm{x}) \).
The loss function was defined as follows:
\begin{align*}
& \mathcal{L} = \left\| B_i(\bm{x}) - \sum\nolimits_j g_j(\bm{x}) b_{ji}(\bm{x}) \right\|_1 s.t. \sum\nolimits_j g_j(\bm{x}) = 1
\label{loss_clearsky_probability}
\end{align*}
Here, \( \bm{x} \) represents the image coordinates.

\begin{figure}[ht]
\vskip 0.1in
\begin{center}
\centerline{\includegraphics[trim=0cm 0cm 0cm 0cm, height=2.5cm]{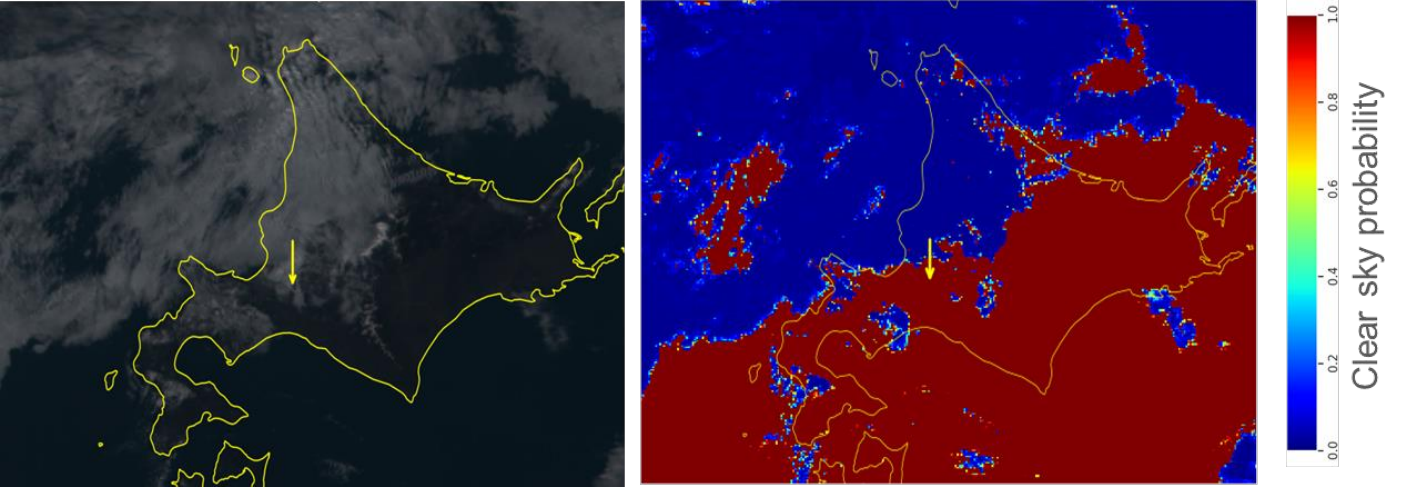}}
\caption{Visible satellite image (left) and distribution of estimated clear-sky probability (right) for the Hokkaido region,
    spanning latitudes 41°N to 46°N and longitudes 138°E to 146°E, on December 3, 2018, at 12:00.
The yellow arrow points to ST1, the ground observation site.
}
\label{clearsky_probability_map2}
\end{center}
\vskip -0.2in
\end{figure}

The process flow and an example of the results are shown in \cref{satellite_segmentation5}.
Here, other information such as outputs from numerical weather predictions and temporal changes in satellite images was referenced to identify the radiation sources for each cluster \( j \).
By aggregating the contribution of radiation sources corresponding to the surface, such as seas, forests, and snow covers, the clear sky probability can be calculated:
\begin{align*}
P_{\text{clr}}(\bm{x}) = \sum\nolimits_{j \in \text{Surface}} g_j(\bm{x})
\end{align*}
In an event shown in \cref{clearsky_probability_map2}, ST1 site is covered with snow under clear-sky.
Although clouds in the vicinity cannot be distinguished from snow covers in visible images,
the clear-sky probability calculated by this sub-model shows a value close to 1, confirming the model's validity.

\section{Results and Conclusions}

In this paper, we demonstrated the effectiveness of utilizing machine learning and multiband data by comparing the developed model against the standard model across cases of thin clouds and snow covers.

\begin{figure}[ht]
\vskip 0.1in
\begin{center}
\centerline{\includegraphics[trim=0cm 0cm 0cm 0cm, height=3cm]{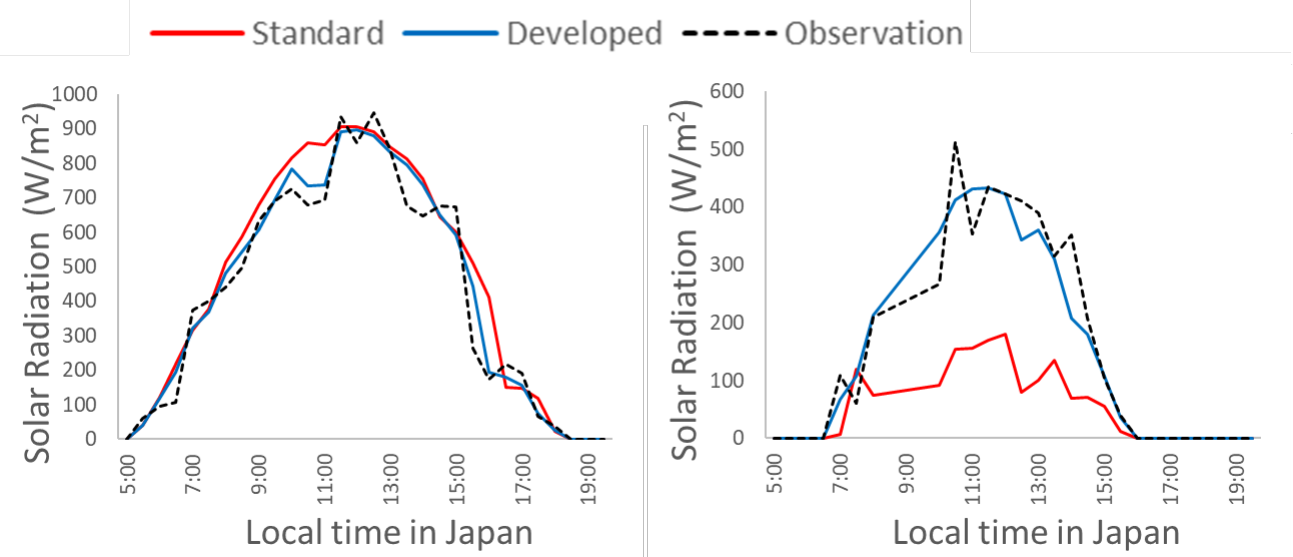}}
\caption{Comparison of solar irradiance time series. Black dotted, red, and blue lines indicate ground observations, the standard model, and the developed model, respectively. Left: ST2, August 15, 2018, at 10:00. Right: ST1, December 3, 2018.}
\label{result_tsukuba_naganuma3}
\end{center}
\vskip -0.2in
\end{figure}

For the thin clouds case, we refer to an event for ST2 on August 15, 2018.
The presence of thin clouds in the area around ST2 reduced solar irradiance, but the standard model overestimated the solar irradiance around 10:00 (see \cref{result_tsukuba_naganuma3}, left).
This overestimation is likely due to surface reflections transmitted through thin clouds being observed in the visible band, leading to an underestimation of cloud thickness.
In contrast, the developed model accurately captured the reduction in solar irradiance caused by the thin clouds.
For this event, when only the visible band was used in the developed model, an overestimation similar to that of the standard model occurred.
This experimental result indicate that by extracting information about the refractive index of ice particles and the vertical structure of upper-level clouds from the near-infrared and infrared band data,
the cloud thickness could be accurately estimated.

For the snow case, we refer to an event for ST1 on December 3, 2018, at 12:00, shown in \cref{clearsky_probability_map2}.
The solar irradiance time series for the day appears in \cref{result_tsukuba_naganuma3} (right).
The standard model cannot distinguish snow cover from clouds, leading to an underestimation of solar irradiance.
However, the developed model accurately estimates solar irradiance.
This accuracy is achieved by correctly identifying snow cover based on the clear-sky probability.
Snow, composed of ice particles and accumulated on the ground, can be distinguished by appropriately combining near-infrared and infrared bands,
although the relationship is inherently complex.
The standard model, which employs a rule-based classification method utilizing several bands, often misidentifies snow as clouds due to its simplistic approach.
In contrast, the developed model uses six bands and leverages machine learning to accurately represent this complexity, thus achieving precise identification of snow and clouds.

To quantitatively compare the standard and developed models, we calculated the Root Mean Square Error (RMSE) for the year 2018.
The calculations used one-minute average values for ground observations and instantaneous values estimated from satellite observation data.
\cref{rmse-table} shows the RMSE for each model by validation site.
The developed model shows an improvement trend at all sites compared to the standard model,
with more than a 10\% improvement effect, particularly when excluding ST5, where localized clouds frequently occur that are not captured by satellites.

These results confirm that the developed model outperforms the standard model.
The main factors for improvement include enhanced representation of thin clouds and convective clouds (not shown in this paper) through the use of multiband data and machine learning,
and the introduction of radiative transfer models and clear-sky probability,
which have improved the generalization performance for various surface conditions including snow covers.
Additionally, the use of CNN has contributed by taking into account the spatial structure of clouds and improving the representation of scattered irradiance.

\begin{table}[t]
\caption{Comparison of RMSE$[\text{W/m}^2]$ between ths standard model and the developed model by validation site.}
\label{rmse-table}
\vskip 0.1in
\begin{center}
\begin{small}
\begin{sc}
\begin{tabular}{lp{0.6cm}p{0.6cm}p{0.6cm}p{0.6cm}p{0.6cm}}
\toprule
Model & ST1 & ST2 & ST3 & ST4 & ST5 \\
\midrule
Standard &106.9&98.2&101.5&111.3&124.8 \\
Developed &86.6&86.1&86.0&94.2&118.6\\
\bottomrule
\end{tabular}
\end{sc}
\end{small}
\end{center}
\vskip -0.1in
\end{table}

This research provides a versatile framework for constructing solar irradiance estimation models applicable to any region,
based on limited ground observation data, not only in Japan but worldwide.
This contributes to the creation of foundational data for meteorological and climatic analyses.

%


\newpage

\section*{Impact Statement}
This paper presents work whose goal is to advance the field of Machine Learning and Earth Science.
There are many potential societal consequences of our work, none which we feel must be specifically highlighted here.

\section*{Acknowledgments}
This study is based on results obtained by the project “development of solar radiation forecasting technologies for short-term forecast of photovoltaic power (Research on short-term forecast of solar radiation)”,
funded by the New Energy and Industrial Technology Development Organization (NEDO).

\bibliographystyle{icml2024}
\bibliography{paper_sasaki}

\begin{thebibliography}{18}
\providecommand{\natexlab}[1]{#1}
\providecommand{\url}[1]{\texttt{#1}}
\expandafter\ifx\csname urlstyle\endcsname\relax
  \providecommand{\doi}[1]{doi: #1}\else
  \providecommand{\doi}{doi: \begingroup \urlstyle{rm}\Url}\fi

\bibitem[Akimoto et~al.(2020)Akimoto, Zhu, Jin, and Aoki]{akimoto2020fast}
Akimoto, N., Zhu, H., Jin, Y., and Aoki, Y.
\newblock Fast soft color segmentation.
\newblock In \emph{Proceedings of the IEEE/CVF conference on computer vision
  and pattern recognition}, pp.\  8277--8286, 2020.

\bibitem[Cornejo-Bueno et~al.(2019)Cornejo-Bueno, Casanova-Mateo, Sanz-Justo,
  and Salcedo-Sanz]{cornejo2019machine}
Cornejo-Bueno, L., Casanova-Mateo, C., Sanz-Justo, J., and Salcedo-Sanz, S.
\newblock Machine learning regressors for solar radiation estimation from
  satellite data.
\newblock \emph{Solar Energy}, 183:\penalty0 768--775, 2019.

\bibitem[Engerer et~al.(2017)Engerer, Bright, and
  Killinger]{engerer2017himawari}
Engerer, N.~A., Bright, J.~M., and Killinger, S.
\newblock Himawari-8 enabled real-time distributed pv simulations for
  distribution networks.
\newblock In \emph{2017 IEEE 44th Photovoltaic Specialist Conference (PVSC)},
  pp.\  1405--1410. IEEE, 2017.

\bibitem[Hale \& Querry(1973)Hale and Querry]{hale1973optical}
Hale, G.~M. and Querry, M.~R.
\newblock Optical constants of water in the 200-nm to 200-$\mu$m wavelength
  region.
\newblock \emph{Applied optics}, 12\penalty0 (3):\penalty0 555--563, 1973.

\bibitem[Hashimoto \& Yoshimoto(2023)Hashimoto and
  Yoshimoto]{hashimoto2023development}
Hashimoto, A. and Yoshimoto, K.
\newblock Development of a short-term solar irradiance forecasting using
  satellite image in combination with numerical weather prediction model.
\newblock \emph{Electrical Engineering in Japan}, 216\penalty0 (3):\penalty0
  e23432, 2023.

\bibitem[Ishii et~al.(2013)Ishii, Otani, Itagaki, and
  Utsunomiya]{ishii2013simplified}
Ishii, T., Otani, K., Itagaki, A., and Utsunomiya, K.
\newblock A simplified methodology for estimating solar spectral influence on
  photovoltaic energy yield using average photon energy.
\newblock \emph{Energy Science \& Engineering}, 1\penalty0 (1):\penalty0
  18--26, 2013.

\bibitem[Jiang et~al.(2019)Jiang, Lu, Qin, Tang, and Yao]{jiang2019deep}
Jiang, H., Lu, N., Qin, J., Tang, W., and Yao, L.
\newblock A deep learning algorithm to estimate hourly global solar radiation
  from geostationary satellite data.
\newblock \emph{Renewable and Sustainable Energy Reviews}, 114:\penalty0
  109327, 2019.

\bibitem[Kondo(1994)]{kondo1994}
Kondo, J.
\newblock Meteorology of water environment.
\newblock pp.\  55--91, 1994.

\bibitem[Liou(2002)]{liou2002introduction}
Liou, K.-N.
\newblock \emph{An introduction to atmospheric radiation}, volume~84.
\newblock Elsevier, 2002.

\bibitem[Mueller et~al.(2009)Mueller, Matsoukas, Gratzki, Behr, and
  Hollmann]{mueller2009cm}
Mueller, R., Matsoukas, C., Gratzki, A., Behr, H., and Hollmann, R.
\newblock The cm-saf operational scheme for the satellite based retrieval of
  solar surface irradiance—a lut based eigenvector hybrid approach.
\newblock \emph{Remote Sensing of Environment}, 113\penalty0 (5):\penalty0
  1012--1024, 2009.

\bibitem[Palacios et~al.(2022)Palacios, Guerrero, and
  Ordo{\~n}ez]{palacios2022machine}
Palacios, L. E.~O., Guerrero, V.~B., and Ordo{\~n}ez, H.
\newblock Machine learning model to predict solar radiation, based on the
  integration of meteorological data and data obtained from satellite images.
\newblock \emph{arXiv preprint arXiv:2204.04313}, 2022.

\bibitem[Ronneberger et~al.(2015)Ronneberger, Fischer, and
  Brox]{ronneberger2015u}
Ronneberger, O., Fischer, P., and Brox, T.
\newblock U-net: Convolutional networks for biomedical image segmentation.
\newblock In \emph{Medical image computing and computer-assisted
  intervention--MICCAI 2015: 18th international conference, Munich, Germany,
  October 5-9, 2015, proceedings, part III 18}, pp.\  234--241. Springer, 2015.

\bibitem[Saito et~al.(2018)Saito, Sasaki, Itagaki, Utsunomiya, and
  Yamaguchi]{saito2018}
Saito, T., Sasaki, K., Itagaki, A., Utsunomiya, K., and Yamaguchi, K.
\newblock Preliminary study on a cloud and snow classification of a solar
  radiation estimation using the meteorological satellite himawari-8 data.
\newblock \emph{The transactions of the Institute of Electrical Engineers of
  Japan. B, A publication of Power and Energy Society}, 138\penalty0
  (6):\penalty0 460--465, 2018.

\bibitem[Shimizu et~al.(2017)Shimizu, Koutarou, and Mikito]{shimizu2017}
Shimizu, A., Koutarou, S., and Mikito, Y.
\newblock Image characteristics of the 16 bands of himawari-8's ahi.
\newblock \emph{Meteorological Satellite Center technical note}, \penalty0
  (62):\penalty0 39--71, 2017.

\bibitem[Warren(1984)]{warren1984optical}
Warren, S.~G.
\newblock Optical constants of ice from the ultraviolet to the microwave.
\newblock \emph{Applied optics}, 23\penalty0 (8):\penalty0 1206--1225, 1984.

\bibitem[Xie et~al.(2016)Xie, Sengupta, and Dudhia]{xie2016fast}
Xie, Y., Sengupta, M., and Dudhia, J.
\newblock A fast all-sky radiation model for solar applications (farms):
  Algorithm and performance evaluation.
\newblock \emph{Solar Energy}, 135:\penalty0 435--445, 2016.

\bibitem[Yeom et~al.(2019)Yeom, Park, Chae, Kim, and Lee]{yeom2019spatial}
Yeom, J.-M., Park, S., Chae, T., Kim, J.-Y., and Lee, C.~S.
\newblock Spatial assessment of solar radiation by machine learning and deep
  neural network models using data provided by the coms mi geostationary
  satellite: A case study in south korea.
\newblock \emph{Sensors}, 19\penalty0 (9):\penalty0 2082, 2019.

\bibitem[Zhang et~al.(2018)Zhang, He, Liang, Wang, and Yu]{zhang2018estimation}
Zhang, Y., He, T., Liang, S., Wang, D., and Yu, Y.
\newblock Estimation of all-sky instantaneous surface incident shortwave
  radiation from moderate resolution imaging spectroradiometer data using
  optimization method.
\newblock \emph{Remote sensing of environment}, 209:\penalty0 468--479, 2018.

\end{thebibliography}

\end{document}